\documentclass[11pt]{article}
\usepackage[margin=3cm]{geometry}
\usepackage{amsmath, amsthm, amssymb, amsfonts}
\usepackage{natbib}
\usepackage{graphicx}
\usepackage[usenames]{color}

\def\b#1{\mbox{\boldmath $#1$}}    

\usepackage{setspace} 

\title{Ecological fallacy and covariates: new insights based on multilevel modelling of individual data}
\author{Michela Gnaldi,\\ Department of Political Sciences, University of Perugia, Perugia, Italy,  michela.gnaldi@unipg.it\\
Venera Tomaselli,\\ Department of Political and Social Sciences, University of Catania, Catania, Italy\\
Antonio Forcina,\\ Department of Economics, University of Perugia, Perugia, Italy}

\begin{document}
\doublespacing
\maketitle
\begin{abstract}
This paper deals with the issue of ecological bias in ecological inference. 
We provide an explicit formulation of the conditions required for the ordinary ecological regression to produce unbiased estimates and argue that, when these conditions are violated, any method of ecological inference is going to produce biased estimates. These findings are clarified and supported by empirical evidence provided by comparing the results of three main ecological inference methods with those of multilevel logistic regression applied to a unique set of individual data on voting behaviour. The main findings of our study have two important implications that apply to all situations where the conditions for no ecological bias are violated: (i) only ecological inference methods that allow to model the effect of covariates have a chance to produce unbiased estimates;
(ii) the set of covariates to be included in the model to remove bias is limited to the marginal proportions. 
Finally, our results suggest that, when the association between two ecological variables is very weak, it is not possible to obtain unbiased estimates even by an appropriate model that accounts for the effect of relevant covariates.
\end{abstract}
{\bf Keywords}: Ecological inference, voting behaviour, logistic regression, multilevel models.

\section{Introduction}
In studies of voting behaviour, an important issue is the shape and strength of association between the choices of voters in a given election and those in a different election. One may also be interested to study how much voters' choices depend on their sex, age, education and social class. However, due to the special nature of electoral data, the true joint distributions between such pairs of variables cannot usually be observed and the only source of information can be extracted either from official electoral data aggregated at the level of local units, such as precincts or polling stations, or from sampling surveys.

Inference based on aggregated data has the advantage, relative to those based on sample surveys, to be inexpensive and to refer to the whole population under investigation. Sample surveys, on the other hand, though based on direct answers given by voters, may be biased when the number of non-respondents is not negligible. Additional bias may be caused by respondents not saying the truth or non remembering what they voted in the last election. An attempt to compare the estimates provided by the two approaches in the study of the association between the choices made in two related elections is presented in \citet{Russo14}. However, because in her survey data the proportions of voters for different parties in the elections seem to be substantially different from the corresponding proportions in the official data, it seems difficult to draw valid conclusions from her study where the data about the true joint distribution are not available. \citet{Liu07} also compared the performance of sample surveys to that of several ecological inference methods for estimating the association between race and propensity to register in a New Orleans mayoral election. He showed that, relative to the proportions of registered voters provided by true data, most ecological inference methods did better than survey data in that specific context.

Since \citet{Robinson}'s seminal paper, it is well known that the association between two variables estimated from data aggregated within geographical units, like polling stations, may be substantially biased relatively to the association that would emerge if data recorded at the individual level were available. The phenomenon, which came to be known as the {\em ecological fallacy}, was used by Robinson as an argument for banning ecological inference from sociological investigations. Nowadays, this recommendation has not, perhaps, many followers due to a better understanding of the conditions which may produce an ecological fallacy and the emergence of more sophisticated methods of ecological inference. In addition, as \citet{Subra09} pointed out, in certain contexts, the degree of association at the individual level may depend on modelling assumptions and thus may not be such an objective quantity as Robinson seemed to believe. An important implication of this result is that, when the results of ecological and individual level studies do not agree, additional investigation may be necessary before concluding that the ecological estimates are inappropriate.

Though in recent years the reputation of ecological inference has grown within the scientific community, there is still substantial disagreement about the relative merits of different ecological inference methods. Because King's methods \citep{Ki99,Ki04} are, perhaps, the most popular, a big bulk of studies focuses on the merits and shortcomings of these methods. For example, \citet{Freed:98} apply King's model to several datasets where individual-level data are available. They find that the King's method gives essentially the same estimates as ecological regression and that both estimates are far from truth. See also \citet{Huds:10} who presents a critical reading of ecological inference techniques, using data coming from New Zealand historic elections, and limited to the 2x2 tables. On a similar ground, \citet{TamCho:98} reports that the point estimates provided by the King's basic model and by ecological regression are indistinguishable, after accounting for the standard errors. Contrary to \citet{Freed:98} \citet{Liu07} claims that  King's extended model with covariates performs best.
As to the extended versions of the King's method, \citet{Freed:98} and \citet{TamCho:98} show that the enhancement of the model due to the inclusion of covariates depends crucially on the specific covariates included in the model; in other words, while the inclusion of certain covariates produces ecological estimates which are closer to those in the individual data, the diagnostics proposed in the King's framework are of little help in the choice of the covariates to be included in the model.

The objective of this paper is a deeper investigation into the ecological bias and, therefore, goes beyond an inspection of the King's methods. First we review and clarify the conditions required for ecological inference to produce unbiased estimates and provide a simple argument which shows that the same conditions are required by any method of ecological inference to avoid bias. This result has two important implication: (i) no simple solution to ecological fallacy exists, but (ii) the bias may be corrected by modelling the effect of a well defined set of covariates. 

To focus our investigation, we concentrate on three ecological inference methods: the so called \citet{Goodman} model, the ordinary least square (OLS) version of King's Multinomial-Dirichlet model \citep{RoJiKiTa} and a revised version of the model proposed by \citet{BrownPayne}.
The latter model may be seen as a refinement of the Goodman model, in that transition probabilities are allowed to vary at random between polling stations and may depend on available covariates.

Empirical support for our findings is provided by the analysis of an extensive set of individual data on voting behaviour from the Democratic Party primary election for the candidate mayor in the city of Palermo, Italy. The individual records contain information on the decision to vote, and on voters' sex and age. In the ecological version, voters are aggregated at the level of polling stations, small local units with an average of about 950 voters. For these data, all three ecological inference methods lead to estimates which are substantially different from those provided by the individual data which, in our context, do not depend on modelling assumptions. The detected biases have an explanation along the lines of \citet{Wake}: the proportions of people who turn out to vote within each group, defined by sex and age, are highly correlated with the relative size of these groups. The fact that the estimates continue to be biases even if we fit ecological inference models where the relative size of such groups are used as covariates is explained by the fact that
the association between ecological variables is very weak, that is, sex and age are very poor predictors of voting decision.

The paper is organised as follows. In Section 2 we recall the main features of ecological data, review some methods of ecological inference and explain why, even when individual data are available, the assessment of the nature of the association may depend on modelling assumptions. In Section 3 the dataset on voting behaviour for the city of Palermo is described.  In Section 4 we analyse the data, clarify why ecological bias is to be expected in this context and why, even the use of appropriate covariates, can not solve the problem. In Section 5 we discuss the main results.
\section{Methodological aspects}
We first introduce the notation and describe the models we will apply in the following.
Let $Y$ denote the set of $C$ options available to voters in a given election and $X$ a discrete variable with $R$ categories which we expect to be associated with $Y$, like, for instance, age groups, sex, social class or the choice made in a previous election.

Suppose we are interested in the association between $X$ and $Y$ within a given area, like a borough divided into a collection of $N$ polling stations. Let $n_{uij}$ denote the number of voters in polling station $u$ with $X=i$ and $Y=j$; in most cases, these records, which we may call {\em individual data}, cannot be observed and we have to rely on the aggregated data:
$x_{ui}$, the number of those who voted $X=i$ in the previous election and $y_{uj}$ the number of those who vote $Y=j$ in the new election.
Let $p_{uij}$ denote the proportion of voters who choose option $j$ among those with $X=i$ in polling station $u$; if $X$ and $Y$ were binary variables, like for instance when $X$ denotes sex and $Y$ is whether one goes to vote or not, the association may be measured by the correlation coefficient or the odds ratio. With $R\times C$ tables, the degree of association is 0 when the probabilities $p_{uij}$ do not change with $i$.
\subsection{Ecological inference methods}
In this paper we focus on three well-known ecological inference methods, by stressing the different assumptions they are based on. Let $\pi_{ij}$ denote the probability that a voter with $X=i$ choose $Y=j$.

The so called Goodman (1953) model is essentially based on two rather strong assumptions according to which (i) the probability that a voter with $X=i$ chooses $Y=j$ does not depend on polling stations and is equal to $\pi_{ij}$ and (ii) voters decide independently from each other. These assumptions imply that the $n_u$ voters in polling station $u$ split among the $C$ options according to a sum of $R$ multinomial distributions. Under these assumptions no ecological fallacy can arise and an unbiased estimate of the $\pi_{ij}$ parameters can be computed by OLS applied to the following set of equations where we write $v_{uj}$ = $y_{uj}/n_u$ and $t_{ui}$ = $x_{ui}/n_u$ to denote the $Y$ and $X$ marginal proportions in each polling stations, respectively.

Because the marginal proportions $t_{ui}$ sum to 1, the proportion for $X=R$, that is $t_{uR}$, may be replaced with $1-\sum_1^{R-1}t_{ui}$
\begin{equation}
v_{uj}=\sum_{i=1}^R t_{ui}\pi_{ij}+\epsilon_{uj}=\pi_{Rj}+\sum_{i=1}^{R-1}(t_{ui}-t_{uR})\pi_{ij} +\epsilon_{uj}.
\label{eq.Goodman}
\end{equation}
Note also that, because the probabilities $\pi_{ij}$ also sum to 1 within each row, the equation for the last column where $j=C$ is redundant.

A collection of methods proposed by Gary King and his co-workers - see, for instance, \cite{Ki97}, \cite{Ki99} and \cite{RoJiKiTa} - have become very popular within certain scientific communities where they are considered to be the most advanced methods of ecological inference, though their merits are debated, see \cite{Liu07} for a review. The well-known basic King model \citep{Ki97, Ki99} incorporates three assumptions. First, the parameters are not constant and their variation can be described by a truncated bivariate normal distribution. Second, there is no "aggregation bias"  that is the parameters are assumed to be uncorrelated with the regressors. Third, the data do not exhibit any spatial autocorrelation. In spite of their sophisticated Bayesian framework, the above assumptions, when translated into models of voting behaviour, are rather simple. For instance, the model described in \cite{RoJiKiTa}, imply that the probability that a voter in polling station $u$ chooses option $j$ equals $\sum_i n_{ui} p_{uij}$. This is equivalent to assume that all voters in polling station $u$ behave as an homogeneous group, irrespective of their value of $X$, an assumption which seems rather unrealistic, especially in the context of voting behaviour when $X$ is the party voted in a previous election occasion; a similar criticism was raised by \citet{GreinerQuinn}.

The model proposed by \citet{BrownPayne} may be seen as a refinement of the Goodman model both in the assumptions on voting behaviour and in the method used to estimate the unknown parameters. More precisely, the model assumes that a voter, who lives in polling station $u$ and had chosen $X=i$, will choose the voting option $Y=j$ with a probability which is no longer the same in all polling stations, but may vary at random as in a Dirichlet distribution whose average, which we may call again $\pi_{ij}$, is common to all polling stations, if covariates are not available. In addition, because parameters are estimated by maximum likelihood on a logistic scale, the estimates are more efficient and lie always between 0 and 1. Recently, \citet{FoGnBr} have proposed a modified version of the above model by assuming that voters who share the same transition probabilities are only those who have the same $X=i$, live in the same local unit and, in addition, belong to the same circle of friends or relatives. This is equivalent to assume that the decisions of voters are correlated only within smaller clusters whose size may vary at random. This model, which we will apply in the following, differs from the Brown and Payne model only for a component in the expression of the covariance matrix of the observations; readers interested in the  technical details, are referred to \citet{FoGnBr}.
\subsection{The ecological fallacy revisited}
In spite of the impact that Robinson's paper had on the scientific community, the true nature of ecological bias is not always well understood. For instance, according to \citet{Russo14}, the ecological fallacy is due to the "incorrect assumption that individual members of a group have the same characteristics as those of the group taken as a whole", a statement which is too vague to be meaningful, unless the Author wanted to say that ecological inference is based on the assumption that transition probabilities, within each local unit, are the same as in the whole town to which they belong.

Though the conditions required to prevent ecological bias in King's models are clearly stated, see for instance \citet {Ki99}, the title of one of his first papers  "A solution to the ecological inference problem"  \citep{Ki97} may have lead to believe that these methods have some kind of intrinsic protection against the ecological fallacy; for instance \citep{Seligson} seems to be convinced that Gary King has advanced "towards solving the ecological inference problem" as long as we have "relatively homogeneous  ecological units", a rather vague statement, though, somehow, closer to the truth.

Perhaps, one of the simplest explanations of the mechanism underlying the ecological fallacy is provided by \citet{Wake} in the case where $X$ and $Y$ are binary variables and there are only two local units (for example, polling station 1 and 2); a numerical example along those lines is given in Table 1. Moving from the first to the second polling station, the proportion of males increases from 0.2 to 0.6 and, at the same time, the overall proportion of those who turn out to vote decreases from 0.7 to 0.4; thus, any method of ecological inference would lead us to conclude that the proportion of those who turn out to vote is smaller among males relative to females. On the other hand, if we look at the joint distribution within each local unit, we see that the proportion of those who go to vote is higher for males relative to females in each polling station.

INSERT TABLE 1 HERE
TABLE 1: Example of ecological fallacy with two polling stations

The apparent paradox has been produced intentionally as follows: when we go from the first to the second polling station the proportions of those who turn out to vote decrease both among males (from 0.15/0.20 to 0.30/0.60) and among females (from 0.55/0.80 to 0.10/0.40) while, at the same time, the marginal proportion of males increases (from 0.2 to 0.6). In other words, the proportions of voters within males and females are not constant and, in addition, their variations are correlated with the proportions in the row totals.

To examine the case where we have a collection of $N$ polling stations with an $R\times C$ table of association, it is convenient to start with the \citet{Goodman} linear regression model. The model assumes that voters with $X=i$ choose option $Y=j$ with a probability $\pi_{ij}$, constant across polling stations, where voters choose independently from one another. These assumptions imply that the $n_{ui}$ voters in polling station $u$ with $X=i$ split among the $C$ options according to a multinomial distribution. It is useful to derive the set of equations for this model in order to obtain an explicit expression for the error term. We may transform the accounting equation, with simple algebra, to obtain:
\begin{equation}
v_{uj} = \sum_{i=1}^R t_{ui}p_{uij} = \sum_1^R t_{ui}\pi_{ij}+\sum_1^R t_{ui}(p_{uij}-\pi_{ij}),
\end{equation}
the second component in the equation above is the  error term, which we may denote by $\epsilon_{uj}$. By using again the fact that $t_{uR}$ = $1-\sum_1^{R-1} t_{ui}$, we may write:
\begin{equation}
v_{uj} = \pi_{Rj}+\sum_{i=1}^{R-1} (t_{ui}-t_{uR})\pi_{ij} +\epsilon_{uj};
\label{eq.Goodman}
\end{equation}
this is the basic equation in Goodman linear regression model.
The condition for the least square estimates to be unbiased, see for example\citet {Wooldridge}, is that $\epsilon_{uj}$ is uncorrelated with the $t_{ui}$ proportions which are the explanatory variables in the regression model; this condition, in turn, implies, for any $i$, that:
\begin{equation}
E\left[(p_{uij}-\pi_{ij}) \mid t_{u1}, \dots , t_{uR}\right] = \b 0.
\label{nocorr}
\end{equation}

We may summarize the above by saying that a necessary and sufficient condition for ecological regression to provide unbiased estimates is that the $p_{uij}$ proportions are uncorrelated with the marginal proportions $t_{ui}$. This means that the condition for the linear regression model to provide unbiased estimates are satisfied even if the $p_{uij}$ vary at random (like in the Brown and Payne model) or depend upon other variables, as long as these other variables are uncorrelated with the set of marginal proportions $t_{ui}$.

Whether or not the above condition is satisfied, when we fit a regression model like Goodman's, the residuals will always be uncorrelated with the marginal row proportions $t_{ui}$, thus, when only ecological data are available, it is not possible to check the condition for no ecological bias. This is possible, instead, when the joint distribution of $X$ and $Y$ is observed in each polling station. In particular, when, like in our case, $Y$ is binary, a logistic regression model can  be fit to test whether the observed proportions $p_{uij}$ depend on the marginal proportions $t_{ui}$.
\subsection{Individual data, multilevel models and the role of covariates}
We now discuss the issues raised by \citet{Subra09}, that is whether individual data, if available, provide always simple and objective estimates of the conditional distributions of $Y$, the voting decision, given $X$, the predictor. Suppose first that no covariates are available and that, though the proportions $t_{ui}$ of voters with $X=i$ vary at random among polling stations, the voting behaviour conditionally on $X$ is the same. In this simple context, we may sum the joint frequencies $n_{uij}$ with respect to polling stations and estimate the conditional probabilities as:
\begin{equation}
\hat \pi_{ij}=\frac{\sum_u n_{uij}}{\sum_u\sum_j n_{uij}}.
\label{eq.Robinson}
\end{equation}
then, these estimates may be compared with those provided by ecological inference. This is essentially what Robinson did with his census data on race and illiteracy: he computed an estimate of the conditional probabilities for being illiterate within whites and blacks. To summarize his results in a more striking way, he also computed the correlation coefficient in the corresponding $2\times 2$ table and compared it with the one based on conditional probabilities estimated by ecological regression with the States as local units.

When the conditional probabilities of choosing $Y=j$ given $X=i$ are not constant, but vary as a logistic function of covariates measured at the level of polling stations, the degree of association will also vary across polling stations. If, like in our case, $Y$ is binary, one could try to fit a logistic model and then estimate the degree of association conditionally on a specific value of the covariates which is of interest, like their average, or compute an estimate of the conditional probabilities for each polling station and the average. Given the non linearity of the logistic model, the second method should be preferred \citep{GreinerQuinn}.

\citet{Subra09} has studied in detail the case where, in addition to covariates, there is also substantial random variation across local units. They fit several different logistic multilevel models to the same set of individual data used by Robinson, the 1930 US Census data. These are individual data by State with adults classified as illiterate or not and as "white born in US", "white not born in US" and "black". Their results indicate that the models which take into account the additional variability across States (their local units) fit much better than the model based on the naive estimates given in (\ref{eq.Robinson}); the more sophisticated models give a substantially different picture of association between race and illiteracy.
\section{The Palermo dataset: collection and preliminary analyses}
On March 4th 2012, the Democratic Party (PD) held its Primary election to choose the candidate for the mayoral election of the city of Palermo, Italy. Four candidates run for that position within the centre-left coalition. The voter turnout, almost 30,00 voters, was higher than in previous primary elections.

The individual dataset used in this paper was constructed as described next. The official electoral rolls of the municipality of Palermo in the  2012 spring listed 564,405 eligible voters grouped into 593 valid polling stations after excluding those based on temporary communities like hospitals and prisons. These records provided gender, date of birth and address, for each eligible voter.

The PD Primary election was held in 31 voting seats, each made of a suitable collection of neighbouring polling stations. Each eligible voter was then listed in the PD records for the Primary election by seat and polling station as `voter' or 'non voter'. The individual data were obtained by matching the two above lists, eligible voter by eligible voter, and then aggregating by polling station and seat.

An overview of the data is given in Table 2: though the number of participants was slightly more than 28 thousands, that is about 5\% of the eligible voters, participation should be compared to the strength of the centre-left coalition, made of PD and a few minor parties. An assessment of their performance may be estimated from the results of the 2012 mayoral election which took place on May 6th. Though by that time the centre-left coalition had split into two different coalitions, overall they obtained over 16\% of the eligible voters and over 18\% in the residential area. Almost 31\% of the potential participants, estimated from the mayoral election, voted at the Primary election with the highest participation rate recorded in the suburban area (over 33\%).

INSERT TABLE 2 HERE
TABLE 2: Voter turnout at the 2012 Primary elections by areas

In the present study, these individual data may be used to check whether the assumptions required to apply  ecological estimation models are, or are not, violated. The data also allow us to assess the effects of covariates within a multilevel framework that takes into account the different sources of variability \citep{Hox:10,Gold:03,Rodr:Gold:95}.

Our data has a three level hierarchical structure: individual voters (as first level units) nested within polling stations (as second level units), and polling stations nested within seats (as third level units). In multilevel settings, covariates may express voters features as well as polling station properties, so that we are dealing with first and second level covariates, respectively \citep{SnijdersBosker:12}. Specifically, in this study the first level covariates employed are sex and age groups; the second level covariates are two measures of partisan leaning and two further measures expressing specific proportions of voters per sex within age groups.

Table 3 provides further insights into the data, with voters classified by sex and age. The distribution of eligible voters by sex is relatively balanced, even if there are  slightly more women (298,773) than men (265,632). Eligible voters are more frequent in the middle age groups (30-65 years).

If we look at the voter turnout within age groups, we note that voter shares do not differ very much between males and females. However, within the 65-75 group, males vote 1.53 times more than females (0.0654 male proportion and 0.0427 female proportion). Besides, in the 75 and over group, males vote 2.23 times more than females (0.0345 for men and 0.0156 for women). If we compare the 45-65 and the 18-25 age groups, we note that the vote share is 1.56 times higher for men (0.0688 against 0.0442) and 1.48 times higher for women (0.0647 against 0.0435).

To sum up, all the proportions of voters by sex and age show that, tough eligible female voters (298,773) are more represented in absolute terms than men (265,632), men turn out to cast a vote more than women.

INSERT TABLE 3 HERE
TABLE 3:  Eligible voters and voters by sex and age

With the aim to fit multilevel models, we employed four covariates measured at the level of polling stations:

\begin{itemize}
\item $pd$, the proportion of voters for the Democratic Party at the municipal election held a month later (3,8\% on average for all the polling stations within total eligible voters);
\item $idv$, the proportion of voters for the {\em Italia dei Valori} Party at the same municipal election (5,1\% on average for all the polling stations within total eligible voters);
\item $mol$, the proportion of males aged between 45 and 74 (45.0\% within male eligible voters);
\item $fol$, the proportion of females aged between 45 and 74 (45.4\% within female eligible voters).
\end{itemize}
\section{The data analysis}
\subsection {Logistic multilevel models}
\begin{figure}[ht]
\centering
 \includegraphics[width=13cm,height=6cm]{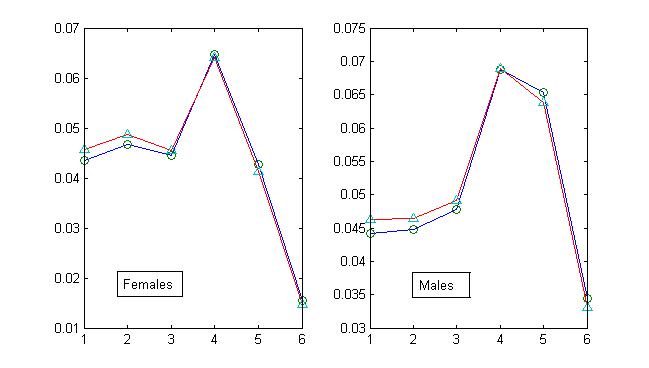}
 \caption{Estimates of the probability to go to vote; age groups on the $x$ axis, females in the left panel, males on the right; $o$ marks raw estimates and {\tiny $\triangle$} those of the multilevel model}
 \label{F1}
\end{figure}

We fitted a set of logistic regression models similar to those in \citet{Subra09} to estimate the size and structure of random variations among polling stations in the propensity to vote and to see how this affects the estimates of the fixed effects. For each polling station, voters are divided according to sex and to 6 age groups as in Table 3. For the multilevel analysis, our data consist of 12 binomial observations: voters and non voters for each of the 6 age groups within each sex, nested within each polling station with polling stations grouped into 31 seats. In this model, there are three sources of variation:
\begin{description}
\item{(i)} binomial within polling stations;
\item{ (ii)} among polling stations within seats with a standard deviation of 0.2311;
\item{(iii)} among seats with a standard deviation of 0.2547.
\end{description}
The above multilevel logistic model  provides estimates of the probability to go to vote within each sex by age group which depends on the covariates and thus are different in each polling station. An overall estimate may be computed by averaging the estimates obtained within each of the 593 polling stations. In Figure \ref{F1} we have plotted the estimates provided by the multilevel model with covariates together with the raw estimates as in (\ref{eq.Robinson}) which ignore covariates and random variation; the two sets of estimates seem to be in rather close agreement.

INSERT FIGURE 1 HERE.
FIGURE 1: Estimates of the probability to go to vote; age groups on the $x$ axis, females in the left panel, males on the right; $o$ marks raw estimates and {\tiny $\triangle$} those of the multilevel model

It can be seen that the three different estimates of voting probabilities are very similar within each sex by age group. Thus, in this context, there is not much ambiguity in determining the pattern of association between sex-age and propensity to vote on the basis of the data at the individual level. In other words, our data provide reliable targets with which to assess the accuracy of the estimates provided by ecological inference.

For the Robinson's data, \citet{Subra09} found that the model (their Model 4) which allows the random effect due to states to be different among their 3 groups by race and birth fitted much better and produced different estimates. In our context, this is equivalent to assume that the random variation due to polling stations and seats is different for each age by sex group. However, because the routine for estimating this model had convergence problems, probably due to the presence of three nested random effects, 12 categories in addition to covariates, we fitted a separate model for each age group.
The estimates of the size of the random effects due to polling stations and seats within each age group are displayed in Table 4.  Random variation across polling stations  within seats, and across seats seem to be of  the same order of magnitude. Though this model is much more complex, the improvement in the fit is rather modest.

INSERT TABLE 4 HERE
TABLE 4:  Standard deviations of the random effects among polling stations within seats and among seats for each age group

\subsection{Ecological inference estimates}
The estimates of voting probabilities provided by three different ecological inference methods without covariates are displayed in Table 5. It is easily seen that they are substantially different from those based on individual data in Table 3: for certain age groups, estimated probabilities are close to 0 while those for other age groups are much too higher than those provided by all methods of estimation based on individual data displayed in Figure 1. Note also that the estimates provided by the King OLS and the revised Brown-Payne methods are rather similar to one another.

INSERT TABLE 5 HERE
TABLE 5:  Ecological inference estimates of the probability of voting by sex and age groups, without covariates

\begin{figure}[ht]
\centering
 \includegraphics[width=13cm,height=6cm]{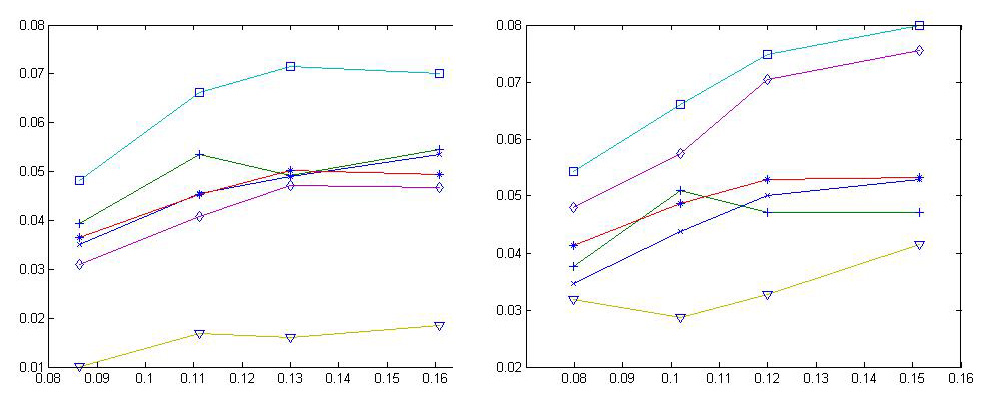}
 \caption{Proportion of voters by age groups as a function of the proportion of eligible votes aged 65-75, Females on the left and males on the right, $+$=18-25, $\times$=25-30, $\ast$=30-45, $\square$=45-65, $\lozenge$=65-75, $\triangledown$=over 75}
 \label{F2}
\end{figure}

Because the results of Section 2.2 tell us that bias should not be present unless the propensity to vote within each group of sex by age is correlated with the relative size of these groups, we may use the individual data to check the size and the direction of these correlations. Consider the quartile regressions plotted in Fig \ref{F2} where polling stations are grouped, as an example, on the basis of the proportion of eligible voters aged 65-75. The average proportion within each quartile is plotted on the $x$ axis against the corresponding average proportion of voters. Most lines have a positive slope indicating that the propensity to vote increase with the proportion of eligible voters aged 65-75.
A similar picture would emerge if we grouped polling stations on the basis of the proportion of voters aged 45-65; instead, if we use as explanatory variable the proportion of voters aged 25-30 or 30-45, regression lines have a roughly negative slope. 

INSERT FIGURE 2 HERE
FIGURE 2: Proportion of voters by age groups as a function of the proportion of eligible votes aged 65-75, Females on the left and males on the right, $+$=18-25, $\times$=25-30, $\ast$=30-45, $\square$=45-65, $\lozenge$=65-75, $\triangledown$=over 75

Further evidence is provided by a more formal procedure based on fitting logistic multilevel models separately for each age group, where the observation at the first level are the number of voters (classified by sex and age category) nested within polling stations nested, again, within seats. As covariates we considered the proportion of eligible voters belonging to each age group separately for males and females, in addition to the $pd$ and $idv$ covariates described above. An informal model selection procedure was used to select which covariates should be included.

Though the proportion of voters aged 45-65 and 65-75 were significant most of the times, when $pd$ and $idv$ were also used, some of the previous covariates appeared to have no longer a significant effect. This could be due to the fact that $pd$ and $idv$ are closely related to the age distribution within each polling station. More precisely, when either $pd$ or $idv$ increases, the proportion of eligible voters in the 18-45 age group decreases while the proportion in the age range from 45 to 75 and over increases. The parameter estimates are displayed in Table 6.

INSERT TABLE 6 HERE:
TABLE 6: Estimated parameters for the multilevel logistic models

The fact that, most of the times, two or more of the covariates are highly significant  indicates that the probabilities to vote for the corresponding age group are strongly correlated with the marginal proportions and thus consistent estimates cannot be obtained by ecological inference in this specific context. The same conclusion is implied by the significance of the $pd$ or $idv$ covariates which are correlated with the marginal distribution of eligible voters.

One may wonder whether ecological estimates that are closer to the truth may be obtained by using appropriate covariates. \citet{Liu07} found that the estimates from King's model improved substantially by including certain covariates. We have fitted a revised Brown and Payne model (see Table 7) where we allow the probability of voting for each sex by age group to depend on the same set of covariates which appeared to have a significant effect in the logistic multilevel models described above. Unfortunately, the resulting estimates are not much better than those given in Table 5.

INSERT TABLE 7 HERE:
TABLE 7: Revised Brown and Payne model: estimates for the probability of voting by sex and age groups with covariates

To gain a better understanding of what is happening, we tried to compare the predictions provided by the revised Brown and Payne model when covariates are included with those obtained from the logistic multilevel models based on individual data. The actual procedure is describe below:
\begin{itemize}
\item for each polling station we compute the overall number of voters predicted by both the Brown and Payne model with covariates and by the set of multilevel models (one for each age group). These predictions were compared with the observed values and the standard deviation of the error for the two methods was computed. This is equal to 15.45 for the Brown-Payne model and 15.94 for the multilevel models, thus, ecological inference with suitable covariates predicts the overall number of voters even more accurately than the multilevel models, which use individual data
\item both estimation methods provide also an estimate of the number of voters classified by age and sex within each polling station and we may compare again these predictions with the observed frequencies and compute the standard deviation of the errors within each sex by age group for the two methods. Here the picture is quite different: the multilevel models provide a substantially better prediction of the number of voters classified by sex and age, as displayed in Figure 2: though for certain groups the two errors are similar, in others ecological inference estimates have a much larger error.
\end{itemize}

INSERT FIGURE 3 HERE
FIGURE 3: Standard deviation in the prediction of the number of voters in each sex by age groups; age groups are on the $x$ axis, females on the left panel; $o$ stands for ecological inference and {\tiny$\triangle$} for individual data

\section{Discussion of main results}
This paper gains a deeper understanding of ecological bias by combining theory and empirical evidence based on the analysis of an extensive set of individual data on voting behaviour from a large sized Italian town. Individual records available for all eligible voters contain information on sex, age and whether they went to vote at the 2012 Democratic Party primary election, in addition to the polling station where the voter is enrolled. Thus, in this study the true joint distributions of eligible voters by sex, age group and the choice to vote or not at the Primary election is known for each polling station.

To uncover the true structure of association from our individual data, we fit a logistic multilevel model with a single random intercept, and a further model with as many random intercepts  as the age groups in our data, as in \citet{Subra09}. Here, voters are 1-level units nested within polling stations (as 2-level units) nested within seats (3-level units). We show that, relatively to our context of study, the pattern of association is unequivocal as the two measures of voting probabilities (averaged across polling stations and for an average polling station, respectively) are very similar one another and similar to the raw proportion estimates, within each sex by age group, too.
The fact that the pattern of association in our data is not model dependent as in the Robinson's data analysed by Subramanian may be due to the fact that, while our data concern a single (though large) town, he was dealing with a big country. In addition, our local units (polling stations) are of very small size while in Subramanian's case they were very large territorial areas.

We apply three ecological inference methods (Goodman's linear regression, a modified version of the Brown and Payne model and the OLS version of King Dirichlet-Multinomial model) and compare ecological estimates with the true proportions in the individual-level data. We find that all the three methods provide estimates which are substantially different from the true proportions. An explanation for this is provided by our results in Section 2 which were, partly, anticipated by \citet{Wake}: the proportion of people who go to vote within each group, by sex and age, are highly correlated with the relative size of these groups. Informal evidence for these correlations is provided by quantile regression. To check formally for this, we fit logistic multilevel model where, among other covariates, we consider the proportion of eligible voters belonging to each sex by age group - that is, the group sizes - and two additional measures of partisan leaning. We ascertain that some covariates are highly significant, so that the probabilities to vote for the corresponding sex by age group are strongly correlated with the marginal proportions obtained by ecological inference. Thus, the conditions to obtain unbiased estimates are not satisfied and this is going to affect any of the three ecological inference methods considered here.

The theory in Section 2 indicates that the only possibility to correct ecological bias is to model the effect of covariates. This is what \citet{Liu07} did in his study, finding that the estimates from the King's model improved substantially by including certain covariates. However, while Liu was searching among all possible covariates, our work indicates how to determine which are the potentially relevant covariates by fitting a logistic multilevel model to the individual data. The interesting result here is that, while the extended version of the Brown and Payne model with covariates does provide a very accurate fit of the total number of voters in each polling station, the estimated proportion of voters by sex and age are not much better than those obtained by the same model without covariates. 

We argue that this is due to the fact that the association between sex and age on one side and voting decision on the other is very weak. When this happens and, in addition, the response variable (voting decision) is binary, the aggregated data provide insufficient information to disentangle the true structure of association from ecological data. In other words, even if we manage to adjust for the sources of ecological fallacy by fitting ecological inference models with covariates, this may not be sufficient to obtain reliable estimates. It may happen, like in our data, that while the probability to vote for a certain age group increases with a covariate, it may decrease for another age group relative to the same covariate in such a way that these two effect partly compensate at the aggregated level. It follows that our ecological inference model may predict very well the variations of the aggregate proportion and, at the same time, provide rather poor estimates of the probability of voting within each age by sex group. 
This last feature goes beyond ecological fallacy. Subsequent analyses show in fact that ecological inference with suitable covariates predicts the overall number of voters in each polling station more accurately than multilevel models based on individual data. However, multilevel models provide a better prediction within each group of the number of voters classified by sex and age. These last results suggest that, even if we manage to adjust for the sources of ecological fallacy by fitting ecological inference models with covariates, this may not be sufficient to obtain reliable estimates. It may happen, like in our data, that while the probability to vote for certain categories increases with a covariate, it may decrease for other categories of the same covariate in such a way that these two effects partly compensate at the aggregated level. It follows that our ecological inference model may predict very well the variations of the aggregate proportion and, at the same time, provide rather poor estimates of the probability of voting within each age by sex group.

Overall, the enhancement of the current state of the art provided by the present study is twofold. 
On one side, we clarify the conditions under which the estimates provided by any  method of ecological inference are likely to be biased.
In so doing, we point out that when voter's choices are modelled without accounting for the effects of covariates, they provide far from truth results.
On the other hand, our results indicate clearly where to look for appropriate covariates which, once taken properly into account, should, remove bias. Except in  special contexts, like in the Palermo data, when the association between the two ecological variables of interest is very weak and the response variable is binary. 
In other terms, even ecological models which are both based on plausible assumptions and extended to account for the effects of significant covariates affecting vote choices, do not necessarily provide reliable estimates. 

\section*{Acknowledgment}{\em{For the Palermo primary data, special thanks to Giovanni Barbagallo, member of the Sicilian Regional Parliament of the Democratic Party.}}


 \newpage

\begin{table}
\caption{\label{Tab:1} Example of ecological fallacy with two polling stations}
\centering
\fbox{
\begin{tabular}{lrrrcrrr}
 & \multicolumn{3}{c}{Polling station 1} & \hspace{2mm} & \multicolumn{3}{c}{Polling station 2} \\ \hline
 & No & Yes & Total & & No & Yes & Total \\ \hline
 F & 0.25 & 0.55 & {\bf 0.80} & & 0.30 & 0.10 & {\bf 0.40}\\
 M & 0.05 & 0.15 & {\bf 0.20} & & 0.30 & 0.30 & {\bf 0.60} \\ \hline
 Tot. & {\bf 0.30} & {\bf 0.70} & {\bf 1.00} & & {\bf 0.60} & {\bf 0.40} & {\bf 1.00}
\end{tabular}}
\end{table}

\begin{table}
\caption{\label{Tab:2}Voter turnout at the 2012 Primary elections by areas}
\centering
\fbox{
\begin{tabular}{lrrrrrrrr} \hline
      &         & Primary  & Primary & Mayoral   & Mayoral & & \\
 Area & Polling  & Eligible & Voters & Centre-Left & Eligible & b/c & a/b \\
     & stations & voters  &  (a) &  votes (b) & voters (c) & $\times 100$ & $\times 100$
 \\ \hline
  City Center & 114 & 107389 & 4406 & 16575 & 107252 & 15.45 & 26.58	\\
  Residential & 180 & 166560 & 9429 & 30768 & 166214 & 18.51 & 30.65	\\
  Suburban    & 299 & 290456 & 14961 & 45216 & 290149 & 15.58 & 33.09	\\
  OVERALL    & 593 & 564405 & 28796 & 92559 & 563615 & 16.42 & 31.11
 \\ \hline
\end{tabular}}
\end{table}

\begin{table}
\caption{\label{Tab:3} Eligible voters and voters by sex and age.}
\centering
\fbox{
\begin{tabular}{lrrrrrrrr} \hline
 & &18-25 & 25-30 & 30-45 & 45-65 & 65-75 & over 75 & Totals \\ \hline
 \multicolumn{9}{c}{Eligible voters by sex and age} \\
Males &  & 32,417 & 22,015 & 70,091 & 89,798 & 29,725 & 21,586 & 265,632\\
Females &  & 30,490 & 21,109 & 73,397 & 99,564 & 36,201 & 38,012 & 298,773\\
   \multicolumn{9}{c}{Voters by sex and age } \\
  Males &  & 1,434 &  986 &	3,351 &	6,179 & 1,945  & 745 & 14,640 \\
Females &  & 1,327 & 986 & 3,268 & 6,438 & 1,545 &	592 & 14,156\\
 \multicolumn{9}{c}{Proportions of voters by sex and age} \\
       Males  & & 0.0442 & 0.0448 & 0.0478 & 0.0688 & 0.0654 & 0.0345 & 265,632\\
Females & & 0.0435 & 0.0467 & 0.0445 & 0.0647 & 0.0427 & 0.0156 & 298,773\\
         \hline
\end{tabular}}
\end{table}

 \newpage

\begin{table}
\caption{\label{Tab:4} Standard deviations of the random effects among polling stations within seats and among seats for each age group}
\centering
\fbox{
\begin{tabular}{lrrrrrrr} \hline
     & 18-25 & 25-30 & 30-45 & 45-65 & 65-75 & over 75 & All together \\ \hline
Poll. stations & 0.2482 & 0.2588 & 0.2628 & 0.2450 & 0.2957 & 0.3339 & 0.2311\\
Seats    & 0.2589 & 0.1906 & 0.2321 & 0.2730 & 0.2782 & 0.2591 & 0.2547
 \\ \hline
\end{tabular}}
\end{table}

\begin{table}
\caption{\label{Tab:5} Ecological inference estimates of the probability of voting by sex and age groups, without covariates}
\centering
\fbox{
\begin{tabular}{lcrrrrrr} \hline
   & Sex & \multicolumn{6}{c}{Age groups}  \\
Method & & 18-25 & 25-30 & 30-45 & 45-65 & 65-75 & over 75 \\ \hline
Goodman & M & 0.000 & 0.000 & 0.000 & 0.123 & 0.147 & 0.000 \\
        & F & 0.000 & 0.000 & 0.000 & 0.124 & 0.000 & 0.028\\
Brown-Payne & M & 0.000 & 0.000 & 0.000 & 0.000 & 0.278 & 0.000 \\
(revised)   & F & 0.000 & 0.000 & 0.000 & 0.148 & 0.000 & 0.152\\
King OLS & M & 0.001 & 0.001 & 0.000 & 0.002 & 0.264 & 0.007  \\
         & F & 0.000 & 0.001 & 0.000 & 0.144 & 0.004 & 0.161 \\ \hline
\end{tabular}}
\end{table}

\begin{table}
\caption{\label{Tab:6} Estimated parameters for the multilevel logistic models
 for the propensity to vote; $F$ is the intercept within females, $M-F$ is the difference in intercept between males and females; $\circ$ = non significant, $\star$ = 5\% significant, $\ast$ = 1\% significant, $\bullet$ = $p$-value smaller than 0.001} .
\centering
\fbox{
\begin{tabular}{lrrrrrr} \hline
Parameters & \multicolumn{6}{c}{Age groups}  \\
 & 18-25 & 25-30 & 30-45 & 45-65 & 65-75 & over 75 \\ \hline
 $F$ & -4.9207$^{\bullet}$ & -4.5096$^{\bullet}$ & -4.4335$^{\bullet}$ & -3.6369$^{\bullet}$& -4.7301$^{\bullet}$ &	-5.0218$^{\bullet}$\\
$pd$ & 13.1413$^{\bullet}$&	14.8040$^{\bullet}$& 8.7711$^{\bullet}$& 12.4372$^{\bullet}$& 7.9891$^{\bullet}$& 7.1119$^{\bullet}$\\
$idv$ &4.5733$^{\star}$&	0.0000$^{\circ}$&	4.2559$^{\ast}$&	5.2272$^{\bullet}$&	4.0019$^{\star}$&	10.3799$^{\bullet}$\\
$P(45-65)$ &2.3815$^{\ast}$&	2.6580$^{\ast}$&	1.6830$^{\ast}$&	0.0000$^{\circ}$&	2.0258$^{\star}$&	0.0000$^{\circ}$\\
$P(65-75)$& 2.3888$^{\ast}$&	0.0000$^{\circ}$&	1.9564$^{\ast}$&	1.3247$^{\ast}$&	2.9426$^{\bullet}$&	0.0000$^{\circ}$\\
$M-F$&0.0256$^{\circ}$&	-0.0570$^{\circ}$& 0.0853$^{\bullet}$&	0.0898$^{\bullet}$&	0.4785$^{\bullet}$&	0.8171$^{\bullet}$
\end{tabular}}
\end{table}

\begin{table}
\caption{\label{Tab:7} Revised Brown and Payne model: estimates for the probability of voting by sex and age groups with covariates}
\centering
\fbox{
\begin{tabular}{crrrrrr} \hline
    Sex & \multicolumn{6}{c}{Age groups}  \\
 & 18-25 & 25-30 & 30-45 & 45-65 & 65-75 & over 75 \\ \hline
 M & 0.000 & 0.001 & 0.000 & 0.241 & 0.110 & 0.053 \\
 F & 0.000 & 0.000 & 0.000 & 0.027 & 0.000 & 0.000
\end{tabular}}
\end{table}

\newpage

\begin{figure}[ht]
\begin{center}
 \includegraphics[width=13cm,height=4cm]{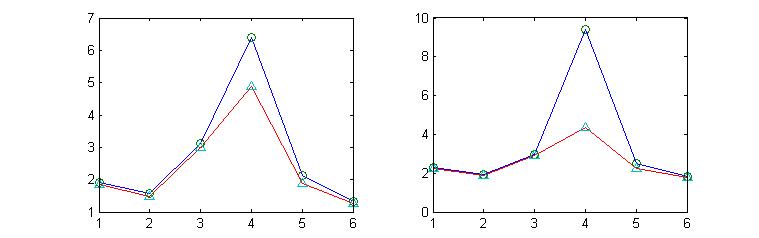}
 \caption{Standard deviation in the prediction of the number of voters in each sex by age groups; age groups are on the $x$ axis, females on the left panel; $o$ stands for ecological inference and {\tiny$\triangle$} for individual data}
\end{center}
\label{F3}
\end{figure}

\end{document}